\documentclass{elsarticle}

\usepackage{amsmath,bm}
\usepackage{graphicx}
\usepackage{amsmath}
\usepackage{amssymb}
\usepackage{color,dsfont}

\makeatletter
\def\ps@pprintTitle{%
 \let\@oddhead\@empty
 \let\@evenhead\@empty
 \def\@oddfoot{\centerline{\thepage}}%
 \let\@evenfoot\@oddfoot}
\makeatother

\begin{document}

\begin{frontmatter}

\title{Long Distance Cavity Entanglement \\by Entanglement Swapping Using
Atomic Momenta}
\author{Sami Ul Haq}
\address{School of Natural Sciences, National University of Sciences and Technology, H-12 Islamabad, Pakistan}
\fntext[myfootnote]{samiulhaq2010@gmail.com}
\author{Aeysha Khalique}
\address{School of Natural Sciences, National University of Sciences and Technology, H-12 Islamabad, Pakistan}
\address{Hefei National Laboratory for Physical Sciences at Microscale and Department of Modern Physics, University of Science and Technology of China, Hefei, Anhui 230026, China}
\address{Shanghai Branch, CAS Center for Excellence and Synergetic Innovation Center in Quantum Information and Quantum Physics, University of Science and Technology of China, Shanghai 201315, China}

\fntext[myfootnote]{aykhalique@yahoo.com}%


\begin{abstract}
We propose a simple technique to generate
entanglement between distant cavities by using entanglement swapping involving atomic momenta. For
the proposed scheme, we have two identical atoms, both initially in their
ground state, each incident on far apart cavities with particular initial
momenta. The two cavities are prepared initially in superposition of zero
and one photon state. First, we interact each atom with a cavity in
dispersive way. The interaction results into atom-field entangled state.
Then we perform EPR state measurement on both atomic momenta state which is an analog of Bell measurement. The EPR state
measurement is designed by passing the atoms through cavity beam splitters which
transfers the atomic momentum state into superposition state. Finally, these
atoms are detected by the detector. After the detection
of the atoms, we can distinguish that cavities in one of the
Bell states. This process leads to two distant cavity fields entanglement.
\end{abstract}

\begin{keyword}
Entanglement swapping, external degrees of freedom,
cavities entanglement, Bragg diffraction, Atomic momentum state, matter-wave
interaction.
\end{keyword}

\end{frontmatter}


\section{Introduction}

Entanglement, a non-local trait of quantum theory, has many applications in
quantum informatics~\cite{EJ96}. The cavity quantum electrodynamics (QED)
techniques are used to generate atom-field, atom-atom and field-field
entanglement~\cite{RBH01}. Entanglement in the atomic external degrees of
freedom using Bragg diffraction is also proposed~\cite{KS03, HS13}. Bragg
diffraction of atomic de-Broglie waves from optical cavity also covers some
aspects of quantum information~\cite{KZ99, QZZ03}.

Entanglement swapping, an important technique of entanglement, entangles two
parties that have never interacted before. Entanglement swapping between two
photons that have never coexisted is demonstrated~\cite{MHS+13}. Bell measurements
are much useful in quantum communication protocols such as teleportation~\cite{BBC+93} and entanglement swapping~\cite{ZZH+93}. Entanglement swapping is used in quantum repeaters~\cite{BBC+93}, in order to overcome the limiting effect of photon loss in long range
quantum communication.

In this paper, we use a simple technique i.e. atomic interferometry for swapping entanglement between atoms and cavities. This way we are able to entangle distant cavities without direct interaction. For the
proposed scheme, we have two cavities which are in superposition state of
zero and one photon. The cavity superposition state is experimentally
demonstrated by Rauschenbeutal et al.~\cite{RNO+00}. First, we interact two
atoms, initially in their ground state having momentum 
$|P_{0}^{i}\rangle$, $i\epsilon \{1, 2\}$ each with a cavity in Bragg
diffraction regime. Bragg scattering allows only one of the two directions of propagation for each atom along the cavity field which are the incident and exactly opposite one. The detuning is large as compared to single photon Rabi frequency and hence atom practically stays in the ground state and the state of the field does not change. We take any order of Bragg scattering in order to allow varying separation between the atoms after the interaction. The non-resonant interaction entangles the
atoms in their external degrees of freedom i.e. in their momentum states
with the cavities. Then these entangled atoms are passed through beam
splitter. For this purpose we use two beam splitters, one for non deflected
atomic momentum state and second for deflected atomic momentum state. The
beam splitter brings the atomic momentum state of these indistinguishable
atoms in superposition state. A cavity in the superposition state of zero
and one photon can be used as a beam splitter~\cite{AIK+09}. At last, after
passing through the beam splitter, these identical atoms are detected. Here,
we use four detectors for four possible momentums splits. The detection process gives us the information that the two cavities are in which Bell state. Thus entanglement between atoms is swapped to that between two far away cavities.

Our paper proceeds as follow: In Sec.~\ref{Bragg Diffraction}, we explain the Bragg diffraction of atom from cavity field and the formation of atom-field entanglement. In Sec.~\ref{Bell Measurement}, we analyze the action of beam splitter which transfers the atomic
momentum component into superposition state. We then briefly explain
the detection process and the final result. Finally we conclude in Sec.~\ref{conclusion} and
give experimental parameters to perform our proposed scheme in the
laboratory.

\section{Bragg atom-field interaction}\label{Bragg Diffraction}

\begin{figure}
\includegraphics[scale=0.5]{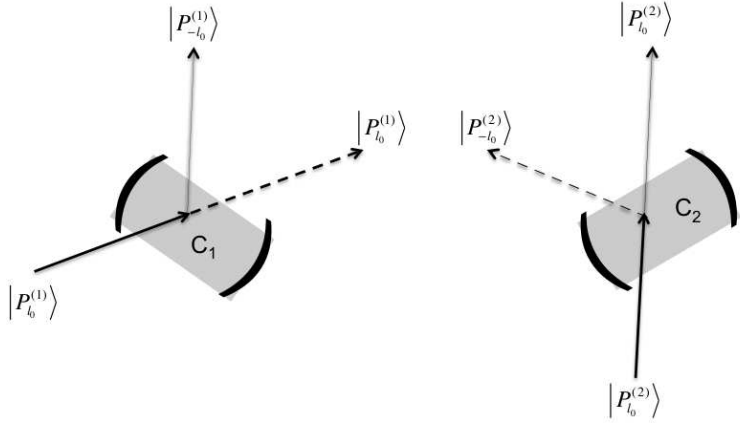}
\caption{We show dispersive interaction of atoms with cavity fields.
The atoms with initial momentum, $|P_{l_0}^{i}\rangle,$ interact
with the cavities which are in superposition of zero and one photon state. The interaction time is set such that when the cavities are in zero photon state, the atoms do not get deflected and have same momentum $|P_{l_0}^{i}\rangle$ as initial one. For one photon state of the cavities, the atoms are deflected and have momentum $|P_{-l_0}^{i}\rangle$.
}
\label{fig:Setup}
\end{figure}

For the proposed scheme, we first entangle two atoms with their respective cavity fields by atom-field interaction in Bragg Regime. For the purpose, we consider two atoms, A$_{1}$ and A$_{2}$, both initially in their ground state, g$%
_{1}$ and g$_{2}$, having transverse momentum state, $|P_{l_0}^{i}\rangle$, where, $i=1, 2$ stands for atoms, A$_{1}$ and A$_{2}$ and $P_{l_0}=\frac{l_0}{2}\hbar k$ with $l_0$ a positive even integer. We
have two cavities, C$_{1}$ and C$_{2},$ which are in superposition state of
zero and one photon i.e. $(|0\rangle +|1\rangle)/\sqrt{2}$ 
\cite{KS03} as shown in Fig. \ref{fig:Setup}. This superposition can be generated by first passing a two level atom in its excited state for half a Rabi cycle through the field. We dispersively interact atom, A$_{1},$
with cavity, C$_{1},$ and atom, A$_{2},$ with cavity, C$_{2}.$ The off-
resonant interaction is followed to avoid decoherence that stems from
spontaneous emission. Large detuning and large interaction time ensure conservation of energy which leads to only two possible directions of scattering for atoms, first the incident one, $P_{l_0}$, and second exactly opposite to the incident transverse momentum direction, $P_{-l_0}$. The off-resonant Bragg diffraction invokes only the
virtual transition among different atomic levels \cite{AIK+09}. The initial state
vector for the system before interaction is 
\begin{equation}
|\Psi (0)\rangle=\frac{1}{2}\sum_{i=1,2}(|0_{i}\rangle
+|1_{i}\rangle )\otimes |g_{i},P_{l_0}^{(i)}\rangle.
\end{equation}
Total Hamiltonian governing this atom-field interaction under the dipole
and rotating wave approximation with atom of mass, $M$, and centre
of mass momentum, $P$, is \cite{KS03}
\begin{equation}
\hat{H}=\frac{\hat{P}^{2}}{2M}+\frac{1}{2}\hbar \omega _{0}\hat{\sigma}
_{z}+\hbar \nu \hat{a}^{\dag }\hat{a}+\hbar g\cos(k\text{ }\hat{x})[\hat{\sigma}_{+}\hat{a%
}+\hat{\sigma}_{-}\hat{a}^{\dag }].
\end{equation}
Here, $\hat{\sigma}_{\pm }$ and $\hat{\sigma}_{Z},$ are the Pauli operators, 
$\hat{x}$ is position operator of atom, $\hat{a}$ ($\hat{a}^{\dag })$ is field
annihilation (creation) operator, $g$ is the vacuum Rabi frequency and $\Delta $ is the detuning between the atomic transition frequency, $\omega _{0}$,
and the field frequency is $\nu$. We follow the large detuning case where
we have no direct atomic transition and it is rare to find the atom in their
excited state. Hence, the system may be governed by following effective Hamiltonian, under the adiabatic approximation as
\begin{equation}
\hat{H}_\text{eff}=\frac{\hat{P}^{2}}{2M}-\frac{\hbar |g|^{2}}{2\Delta }%
\hat{n}\text{ }\hat{\sigma}_{-}\hat{\sigma}_{+}(\cos2k\hat{x}+1).
\end{equation}
The state of each $i$th atom-field pair at any time $t$ is given as
\begin{equation}
|\Psi(t)\rangle=\frac{1}{\sqrt{2}}\sum_{l=-m}^{m}\left(C_{0,\tilde P_l}|0,g_i,\tilde P_l^{(i)}\rangle+C_{1,\tilde P_l}|1,g_i,\tilde P_l^{(i)}\rangle\right),
\end{equation}
where, $m$ is the total number of the orders of deflections and $\tilde P_l=P_{l_0}+l\hbar k$, $l$ being an even integer. Time evolution of the state vector is given by Schrodinger equation
\begin{equation}
i\hbar\frac{\partial |\Psi(t)\rangle}{\partial t}=H_\text{eff}|\Psi(t)\rangle
\end{equation}
We have 
\begin{equation}
\cos2k\hat x|\tilde P_l\rangle\backsim|\tilde P_{(l+2)}\rangle+|\tilde P_{(l-2)}\rangle
\end{equation}
and we drop the unchanged atomic ground state vector $|g_i\rangle$. Under condition of Bragg scattering with only two possible directions of deflection $l=0$ with $\tilde P_0=P_{l_0}$ and $l=-l_0$ with $\tilde P_{-l_0}=P_{-l_0}$. Thus we obtain the state of each $i$th atom-field pair after interaction as \cite{KS03,MS92}
\begin{equation}
|\Psi (t)\rangle =\frac{1}{\sqrt{2}}\left(
|0_{i},P_{l_0}^{(i)}\rangle +C_{1,l_0}(t)|1_{i},P_{l_0}^{(i)}\rangle +C_{1,-l_0}(t)|1_{i},P_{-l_0}^{(i)}\rangle\right) 
\end{equation}
where, $C_{n,\pm l_0}$ is the probability amplitude of the atom exiting with momentum $P_{+l_0}$ or $P_{-l_0}$ when there are $n$ photons in the field and is given as
\begin{equation}
C_{n,\pm l_0}(t)=e^{-iA_nt}\left[C_{n,\pm l_0} (0)\cos \left(\frac{1}{2}B_nt\right)+iC_{n,\mp l_0} (0)\sin \left(\frac{1}{2}B_nt\right)\right]
\end{equation}
where,
\begin{displaymath}
A_n \equiv \left\{ \begin{array}{ll}
-\frac{\left(|g|^2n/{4\Delta}\right )^2}{\omega_{\textrm{rec}}(l_0-2)(2)}
& \textrm{for $l_0 \neq 2$}\\
 0 & \textrm{for $l_0 =2$}
  \end{array} \right.
\end{displaymath}
and 
\begin{displaymath}
|B_n| \equiv \left\{ \begin{array}{ll}
\frac{\left(|g|^2n/{2\Delta}\right)^{l_0/2}}{\left(2\omega_{\textrm{rec}}\right)^{l_0/2-1}\left[\left(l_0-2\right)\left(l_0-4\right)\dots4.2\right]}& \textrm{for $l_0\ne 2$}\\
|g|^2n/{2\Delta}&\textrm{for $l_0=2$}
\end{array} \right.
\end{displaymath}
Initially both atoms are sent with momentum $P_{l_0}$, so probability of finding the exiting atom in either directions flips as a cosine function of interaction time. We adjust the interaction time of atoms with fields to ensure that if there
is one photon in the fields, the atoms definitely get deflected. The adjusted time is thus 
$t=r\pi/|B_n|$, where $r$ is an odd integer. For first order Bragg scattering, this time simplifies to 
$t=\frac{2r\pi \Delta }{|g|^2}$. The wave function of the two atom-field pairs is
\begin{equation}
\label{eq:psit1}
|\Psi (t)\rangle =\left[ \frac{1}{\sqrt{2}}\left(
|0_{1},P_{l_0}^{(1)}\rangle +ie^{-i\phi}|1_{1},P_{-l_0}^{(1)}\rangle \right) \right]
\otimes \left[ \frac{1}{\sqrt{2}}\left( |0_{2},P_{l_0}^{(2)}\rangle
+ie^{-i\phi}|1_{2},P_{-l_0}^{(2)}\rangle \right) \right],
\end{equation}
where, $\phi=r\pi A_1/B_1$. The atoms in their external degrees of freedom becomes entangled with their respective cavity fields. The combined state of the system can be written as
\begin{align}
|\Psi (t)\rangle=\frac{1}{2}( &|0_{1},0_{2},%
P_{l_0}^{(1)},P_{l_0}^{(2)}\rangle +ie^{-i\phi}|0_{1},1_{2},
P_{l_0}^{(1)},P_{-l_0}^{(2)}\rangle +ie^{-i\phi}|1_{1},0_{2},P_{-l_0}^{(1)},P_{l_0}^{(2)}\rangle\nonumber\\
& -e^{-i2\phi}|1_{1},1_{2},P_{-l_0}^{(1)},P_{-l_0}^{(2)}\rangle).
\end{align}
After adding and subtracting some terms and rearranging, we have
\begin{align}
\label{eq:psit2}
|\Psi (t)\rangle=&\frac{1}{4}\left(|P_{l_0}^{(1)},%
P_{l_0}^{(2)}\rangle +e^{-i2\phi}|P_{-l_0}^{(1)},P_{-l_0}^{(2)}\rangle \right) \left(
|00\rangle -|11\rangle \right)  \nonumber\\
&+\frac{1}{4}\left( |P_{l_0}^{(1)},P_{l_0}^{(2)}\rangle
-e^{-2i\phi}|P_{-l_0}^{(1)},P_{-l_0}^{(2)}\rangle \right) \left( |00\rangle
+|11\rangle \right) \nonumber \\
&+\frac{1}{4}ie^{-i\phi}\left( |P_{l_0}^{(1)},P_{-l_0}^{(2)}\rangle
+|P_{-l_0}^{(1)},\text{ }P_{l_0}^{(2)}\rangle \right) \left( |10\rangle
+|01\rangle \right)  \nonumber\\
&+\frac{1}{4}ie^{-i\phi}\left( |P_{l_0}^{(1)},P_{-l_0}^{(2)}\rangle -|P_{-l_0}^{(1)},%
P_{l_0}^{(2)}\rangle \right) \left( |10\rangle -|01\rangle \right) .%
\end{align}%
Here, we entangle the cavities and atoms in four EPR states, separately,
and all the four states are also entangled with each other. Measurement of
atoms in one of the EPR states, projects the two cavities into a
corresponding Bell state. We discuss this process in next section.

\section{EPR State Measurement on Atomic Momenta EPR States}\label{Bell Measurement}
\begin{figure}
\includegraphics[scale=0.5]{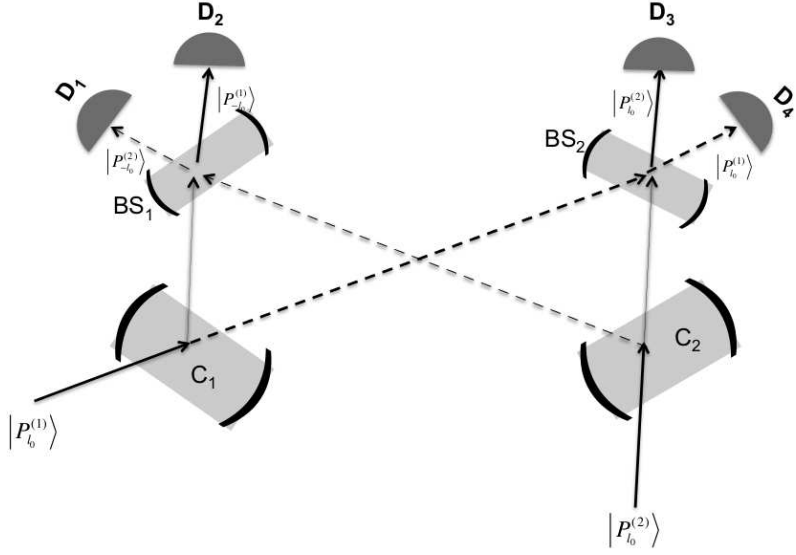}
\caption{For EPR state measurement on atomic momenta, the atoms are passed through the beam splitter BS$_{1}$ and BS$_{2}$
and are finally detected by the detector D$_{1},$ D$_{2},$ D$_{3}$ and D$_{4}.$
The undeflected components pass through beam splitter BS$_{1}$ and deflected
components interact with BS$_{2}.$ The beam splitters are cavities prepared in
superposition of zero and one photon. The two cavities $C_1$ and $C_2$ get entangled after detection of the atomic momenta
states.
}
\label{fig:BellMeasure}
\end{figure}

The scheme proposed in this paper is to develop entanglement between two far away
cavities. EPR state measurement on atomic momentum states collapses the field states into one of the four EPR states. For EPR
measurement on atomic momenta we pass these entangled atoms through beam
splitters. We have two beam splitter, BS$_{1}$ and BS$_{2}.$ The beam splitters are cavities prepared in superposition of zero and one photon. The atomic
momentum components, $|P_{l_0}^{1}\rangle $ and $P_{l_0}^{2}\rangle$, pass through beam splitter, BS$_{2},$ and
momentum components, $|P_{-l_0}^{1}\rangle$ and $|P_{-l_0}^{2}\rangle,$ pass through beam splitter BS$_{1}$ as shown in
Fig.~\ref{fig:BellMeasure}. Mirror can be used to deflect atoms to desired cavities. Here, the two atoms are indistinguishable. The dispersive interaction of
atoms with cavity beam splitter for an interaction time, $t = \frac{2\pi
\Delta' }{|g'| ^{2}}$ \cite{IKS+12}, transfers the atomic momentum states into superposition
state. Here, $\Delta'$ is atom-beam splitter field detuning and $g'$ is vacuum Rabi
frequency. The atoms or beam splitting cavities can be oriented such that atoms undergo first order Bragg scattering. The beam splitter action is performed as follows
\begin{align}
|P_{l_0}^{(1)}\rangle &\longrightarrow|P_{l_0}^{(1)}\rangle +i%
|P_{l_0}^{(2)}\rangle ,\nonumber \\
|P_{l_0}^{(2)}\rangle &\longrightarrow i|P_{l_0}^{(1)}\rangle
+|P_{l_0}^{(2)}\rangle ,\nonumber \\
|P_{-l_0}^{(1)}\rangle &\longrightarrow|P_{-l_0}^{(1)}\rangle +i%
|P_{-l_0}^{(2)}\rangle , \nonumber\\
|P_{-l_0}^{(2)}\rangle &\longrightarrow\dot{\imath}|P_{-l_0}^{(1)}\rangle
+|P_{-l_0}^{(2)}\rangle.
\end{align}
When the beam splitter action is performed, the first factor of the first
term of Eq.~(\ref{eq:psit2}) becomes

\begin{align}
|P_{l_0}^{(1)},P_{l_0}^{(2)}\rangle +e^{-i2\phi}|P_{-l_0}^{(1)},P_{-l_0}^{(2)}\rangle \longrightarrow &\left( |P_{l_0}^{(1)}\rangle +i|P_{l_0}^{(2)}\rangle ,i|P_{l_0}^{(1)}\rangle +|P_{l_0}^{(2)}\rangle
\right)\nonumber\\
&+e^{-i2\phi}\left( |P_{-l_0}^{(1)}\rangle+i|P_{-l_0}^{(2)}\rangle ,i|P_{-l_0}^{(1)}\rangle +|P_{-l_0}^{(2)}\rangle \right)\nonumber\\
=&i(|P_{l_0}^{(1)},P_{l_0}^{(1)}\rangle +|P_{l_0}^{(2)},P_{l_0}^{(2)}\rangle \nonumber\\
&+e^{-i2\phi}|P_{-l_0}^{(1)},P_{-l_0}^{(1)}\rangle
+e^{-i2\phi}|P_{-l_0}^{(2)},P_{-l_0}^{(2)}\rangle).
\label{Eq:Bell1}
\end{align}
Similarly, the first factor of the second term in Eq.~(\ref{eq:psit2}) after the action
of Beam splitter is

\begin{align}
|P_{l_0}^{(1)},P_{l_0}^{(2)}\rangle -e^{-i2\phi}|P_{-l_0}^{(1)},\text{ }P_{-l_0}^{(2)}\rangle
\longrightarrow&\left( |P_{l_0}^{(1)}\rangle +i|P_{l_0}^{(2)}\rangle ,%
i|P_{l_0}^{(1)}\rangle +|P_{l_0}^{(2)}\rangle \right)\nonumber\\
&-e^{-i2\phi}\left(
|P_{-l_0}^{(1)}\rangle +i|P_{-l_0}^{(2)}\rangle ,i%
|P_{-l_0}^{(1)}\rangle +|P_{-l_0}^{(2)}\rangle \right)\nonumber\\
=&i( |P_{l_0}^{(1)},P_{l_0}^{(1)}\rangle +|P_{l_0}^{(2)},P_{l_0}^{(2)}\rangle \nonumber\\
&-e^{-i2\phi}|P_{-l_0}^{(1)},P_{-l_0}^{(1)}\rangle
-e^{-i2\phi}|P_{-l_0}^{(2)},P_{-l_0}^{(2)}\rangle),
\label{Eq:Bell2}
\end{align}
 and that of the third term of Eq.~(\ref{eq:psit2}) is
\begin{align}
|P_{l_0}^{(1)},P_{-l_0}^{(2)}\rangle -|P_{-l_0}^{(1)},\text{ }%
P_{l_0}^{(2)}\rangle \longrightarrow& \left( |P_{l_0}^{(1)}\rangle +i%
|P_{l_0}^{(2)}\rangle ,i|P_{-l_0}^{(1)}\rangle +|P_{-l_0}^{(2)}\rangle \right)\nonumber\\
&-\left( |P_{-l_0}^{(1)}\rangle +i|P_{-l_0}^{(2)}\rangle ,i%
|P_{l_0}^{(1)}\rangle +|P_{l_0}^{(2)}\rangle \right) \nonumber\\
=&2\left( |P_{l_0}^{(1)},P_{-l_0}^{(2)}\rangle
-|P_{-l_0}^{(1)},P_{l_0}^{(2)}\rangle \right),
\label{Eq:Bell3}
\end{align}%
and the same for the fourth term of Eq.~(\ref{eq:psit2}) transform as
\begin{align}
|P_{l_0}^{(1)},P_{-l_0}^{(2)}\rangle +|P_{-l_0}^{(1)},%
P_{l_0}^{(2)}\rangle \longrightarrow&\left( |P_{l_0}^{(1)}\rangle +i%
|P_{l_0}^{(2)}\rangle ,i|P_{-l_0}^{(1)}\rangle +|P_{-l_0}^{(2)}\rangle \right)\nonumber\\
&+\left( |P_{-l_0}^{(1)}\rangle +\dot{\imath}|P_{-l_0}^{(2)}\rangle ,\dot{\imath}%
|P_{l_0}^{(1)}\rangle +|P_{l_0}^{(2)}\rangle \right) \nonumber\\
=&2i\left(|P_{l_0}^{(1)},P_{-l_0}^{(1)}\rangle
-|P_{l_0}^{(2)},P_{-l_0}^{(2)}\rangle \right) .
\label{Eq:Bell4}
\end{align}%
The interaction time of atoms with cavity fields, acting as beam
splitter, can be controlled by using velocity selector.

For the detection of direction of atomic momentum component we use four detectors, D$_{1}$, D$_{2}$, D$_{3}$ and D$_{4}$ \cite{IKS08}. 
The detectors are placed in the spatial paths of the atoms in different directions of propagation of atoms, which correspond to different momenta. Click in the detector corresponds to presence of atom in that direction and hence with that particular momentum. Recently, detectors have been built which can efficiently detect fast moving Rydberg atoms \cite{TKN+09}. A click in detector D$_1$ corresponds to momentum direction $|p_{-l_0}^{(2)}\rangle$, a click in detector D$_2$ corresponds to momentum direction $|p_{-l_0}^{(1)}\rangle$, a click in detector D$_3$ corresponds to momentum direction $|p_{l_0}^{(2)}\rangle$, and a click in detector D$_4$ corresponds to momentum direction $|p_{l_0}^{(1)}\rangle$ as shown in Fig.~\ref{fig:BellMeasure}. Combining Eqs.~(\ref{Eq:Bell1})-(\ref{Eq:Bell4}) with Eq.~(\ref{eq:psit2}), the
state of the cavities for different detector clicks are as given in Table~\ref{Table}. 
\begin{table}[h]
\begin{tabular}{|l|l|l|}
\hline
 Cavities' State&Detectors' click \\ \hline
$|\phi^+\rangle=\frac{1}{\sqrt{2}}\left(|00\rangle +|11\rangle\right) $ & 2 Atoms in either D$_{1}$, D$_{2}$ , D$_{3}$ , or D$_{4}.$ \\ \hline
$|\phi^-\rangle=\frac{1}{\sqrt{2}}\left(|00\rangle -|11\rangle\rangle\right) $ & 2 Atoms in either D$_{1}$, D$_{2}$ , D$_{3}$ , or D$_{4}$. \\ \hline
$|\psi^+\rangle=\frac{1}{\sqrt{2}}\left(|10\rangle +|01\rangle\rangle\right)$ & Coincidence between D$_{1}$
and D$_{4}$ or D$_{2}$ and D$_{3}$.\\ \hline
$|\psi^-\rangle=\frac{1}{\sqrt{2}}\left(|10\rangle -|01\rangle\rangle\right)$ & Coincidence between D$_{2}$
and D$_{4}$ or D$_{1}$ and D$_{3}.$ \\ \hline
\end{tabular}
\caption{The entangled state of the two cavities corresponding to different clicks in the four detectors.}
\label{Table}
\end{table}
Here, from various combination of clicks on detectors, we can distinguish between $|\psi ^{+}\rangle $, $|\psi ^{-}\rangle ,$ and $|\phi ^{+}\rangle $ or $|\phi ^{-}\rangle$. Hence after the detection of atoms we can tell that cavities are in which
Bell state. It must be noted as with linear optics Bell state measurement, this process will only be able to distinguish between states $|\psi^+\rangle$, $|\psi^-\rangle$ and $\{|\phi^+\rangle,|\phi^-\rangle\}$. States $|\phi^+\rangle$ and $|\phi^-\rangle\ $ cannot be distinguished. Deterministic optical Bell measurement schemes rely either on non-linear interactions which are highly inefficient in practice \cite{KKS01}, or using ancilla entangled photons \cite{G11} which require large interferometers to combine the signal and ancilla modes and give near deterministic Bell measurement with asymptotically large ancilla states. Recently single-mode squeezers together with beam splitters have been proposed to give up to 64.3\% success probability \cite{ZvL13}. The EPR state analogs of these near deterministic Bell measurements require further study.

\section{conclusion}\label{conclusion}
We have proposed a scheme for entangling long distance cavities by
entanglement swapping. For this purpose the external degrees of freedom of
atomic momenta are first entangled with cavity fields. We then propose a
method to perform Bell measurement on atomic momenta states which in turn
swaps entanglement to cavity fields. The Bell measurement process involves
additional cavity fields which act as a beam splitters for atomic momenta.

The Scheme that we have proposed for cavities entanglement in Bell states,
possesses stronger non locality. The microwave regime cavity QED have life
time up to seconds and high fidelity can be achieved as proposed by Khosa et.al. \cite{KIZ04}. They consider the
passage of 15-20 helium atoms under the first order Bragg diffraction
through a quantized cavity field. 
We perform quantum
measurement on atomic momentum state of neutral atoms which are easy to
handle and manipulate than a photonic flying qubit. The external atomic
momentum states are more useful states against decoherence \cite{Ball08}. Our
scheme is experimentally feasible as Bragg Scattering of atoms in optical
regime has already been demonstrated by G. Rempe and his co-worker at $\lambda=780$~nm for $^{85}$Rb atoms \cite{KSZ+99}.

\section*{acknowledgments}
We acknowledge valuable discussions with Barry C. Sanders and A.K. greatly acknowledges financial support from the 1000 Talent Program of China.

\section*{References}
\bibliography{References1}
\bibliographystyle{elsarticle-num}

\end{document}